\documentclass{elsart}

\usepackage{graphicx,amssymb}

\begin{document}

\begin{frontmatter}



\title{Investigations of fast neutron production 
       by 190 GeV/c muon interactions on different targets}


\author[neuch]{V. Chazal},
\author[usa]{F. Boehm},
\author[usa]{B. Cook\thanksref{now1}},
\thanks[now1]{Present address: Jet Propulsion Laboratory, Pasadena CA
91109, U.S.A.}
\author[usa]{H. Henrikson},
\author[neuch]{G. Jonkmans},
\author[neuch]{A. Paic\thanksref{now2}},
\thanks[now2]{Present address: Monitor Company, 12 Place Vend\^ome, 75001
Paris, France}
\author[usa]{N. Mascarenhas\thanksref{now3}},
\thanks[now3]{Present address: Foveon Inc., 3565 Monroe St., Santa Clara CA 95051, U.S.A.}
\author[usa]{P. Vogel},
\author[neuch]{J.-L. Vuilleumier  \corauthref{cor}}
\corauth[cor]{corresponding author.}
\ead{jean-luc.vuilleumier@unine.ch}
\address[neuch]{Institut de Physique, Universit\'e de Neuch\^atel,
2000 Neuch\^atel, Switzerland}
\address[usa]{California Institute of Technology, Pasadena CA 91125, U.S.A.}

\begin{abstract}
        The production of fast neutrons (1 MeV - 1 GeV) in high energy
        muon-nucleus interactions is poorly understood, yet it is fundamental
        to the understanding of the background in many underground
        experiments. The aim of the present experiment (CERN NA55) 
        was to measure
        spallation neutrons produced by 190 GeV/c muons scattering on
        carbon, copper and lead targets. 
        We have investigated the energy spectrum and
        angular distribution of spallation neutrons, and we report the
        result of our measurement of the neutron production
        differential cross section. 
\end{abstract}

\end{frontmatter}

\section{Introduction}
        Neutrons are an important source of background in many low
        rate underground experiments. Detectors such as 
        the Palo Verde detector~\cite{PV}, the \linebreak Kamiokande
        detector~\cite{Kamioka}, the CDMS detector \cite{CDMS}, the Gallex
        detector~\cite{gallex}, 
        and most of the other underground experiments are
        concerned with neutron background. It is therefore important
        to understand the different neutron production mechanisms in the
        environment of such experiments. 

        There are two sources of
        neutrons in underground laboratories. The first one
        is the natural radioactivity such as ($\alpha$,n) reactions and
        spontaneous fission, due to uranium and thorium traces in the
        rock or in the materials used in the experiments themselves. The
        second source is the interaction of atmospheric muons, produced
        by cosmic rays, with matter, such as the surrounding rock, 
        the different shieldings or the
        detector itself. Propagating through matter,
        muons lose energy
        mainly continuously by ionization. In addition, they lose energy
        in discrete bursts along their trajectory by bremsstrahlung,
        direct pair production and nuclear interactions~\cite{muons}. 
        The goal of the CERN NA55 experiment was to study
        neutron production in muon-nucleus interactions, 
        which at present is poorly understood. In this
        paper we report the result of our measurement of the
        cross section for fast neutron production by interaction of high
        energy muons off carbon, copper and lead nuclei. 
        Specifically, we investigated the energy spectrum and
        angular distribution of spallation neutrons produced 
        by the inelastic
        scattering of 190 GeV/c muons on each target, via the
        time-of-flight method.
\section{Interaction of high energy muons with matter}
        The total neutron yield measured at different 
        depths has been reported in references~\cite{PV}
        and~\cite{mu1,LVD_n}. 
        However, very little is
	known about the energy and angular distributions of the
	neutrons, even though the neutron energy spectrum 
	in a scintillator has been
	determined in \cite{LVD_n} for neutrons which penetrated at 
	least 1 m from the muon track.
	Moreover, one expects that many neutrons are
	produced in secondary processes following the primary
	muon-nucleus interaction and it is difficult to separate the
	two effects in the inclusive measurements performed
	underground. 

        To describe the process theoretically, one
	assumes that the
	muon scattering occurs primarily at very small angles. Such a
	process can be viewed as produced by a beam of nearly real
	equivalent photons~\cite{delorme}. Even with this drastic
	simplification, it is difficult to obtain
	reliable estimates for actual yields and spectra of the
	neutrons. The difficulty lies in the estimation of
	the equivalent photon flux, as well as in the shortage of
	experimental nuclear photo-disintegration data for high
	energy photons.
\section{Experimental apparatus}
\subsection{Experimental set-up}
        The layout of our NA55 experiment is similar to a CERN LEAR
	experiment~\cite{expcern1} in which neutrons and pions were
	measured following anti-protons stopping in the
	target. Another group, E665 at Fermi-lab~\cite{expcern2},
        measured neutrons produced by deep-inelastic scattering
	of 470 GeV/c muons. However, the latter experiment measured
	neutrons in a small energy range from 1 to 10 MeV, with high-energy
	deep inelastically scattered muons from nuclear targets. In
	the frame of the CERN NA54 experiment, measurements
	were done on a carbon
	target to study the production of radioactive isotopes in
	scintillation detectors by 100 and 190 GeV muons and their secondary
	shower particles~\cite{NA54}.

\begin{figure}
\begin{center}
\includegraphics*[width=15cm]{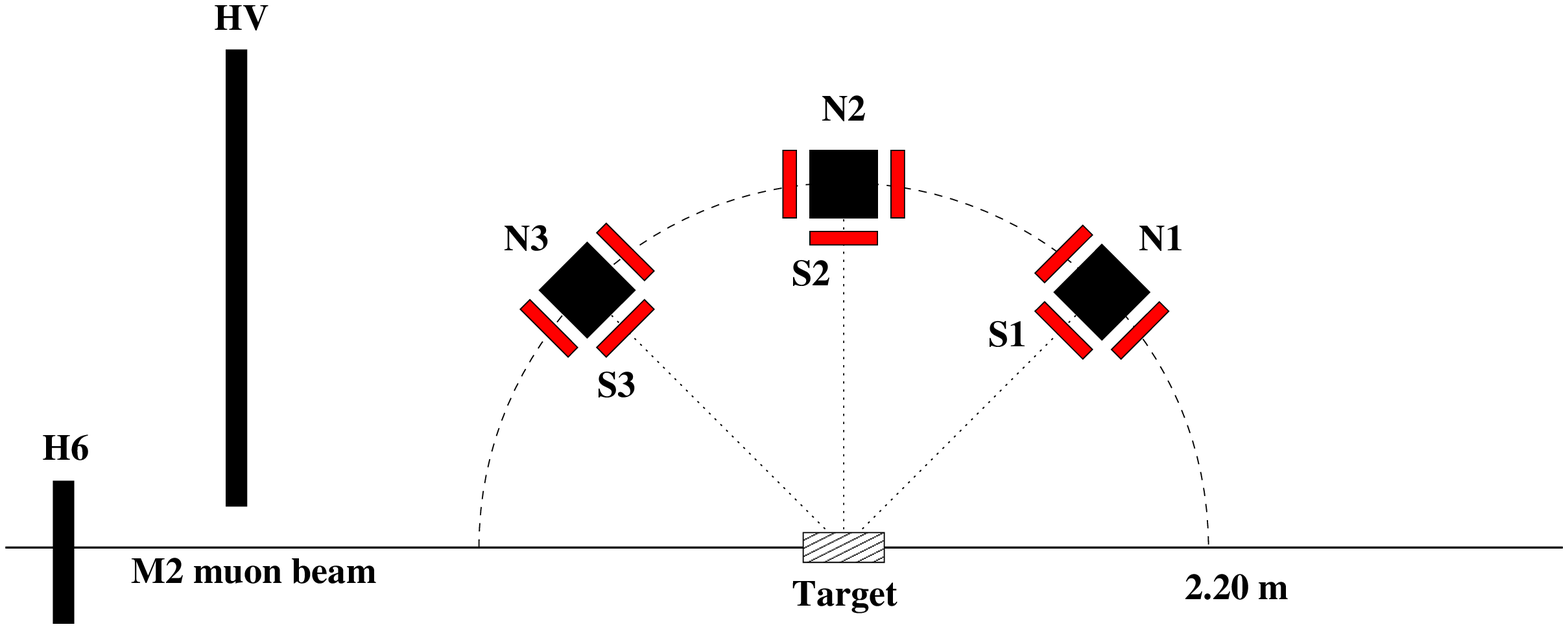}
\vskip 1cm
\caption{\label{setup}NA55 detector layout. N: Bicron 501A liquid
scintillator detectors with Pulse Shape Discrimination. S: thin
plastic scintillator counters for charged particle identification, in
front and around each N detector. H6: NA47 segmented high rate beam
counter. HV: beam muon halo veto, thin plastic scintillator. Arbitrary
figure scale.}
\end{center}
\end{figure}

        The M2 muon beam
	at the CERN-SPS was well suited for our
	investigation. At 190 GeV/c, the energy of the SPS muons is
	similar to the mean energy of cosmic-ray muons at many
	underground experimental sites. The beam transverse size is
	2.2 cm (FMWH). Our experiment was performed
	with a cylindrical target of 75 cm length and 8 cm
	diameter for the graphite, 25 cm length and 10 cm diameter for
	the copper, 10 cm length and 20 cm diameter for the lead
	target. Three neutron detectors (see section 3.2) 
        were placed at 45 (N1), 90 (N2) and
	135 (N3) degrees relative to the muon beam, for a
	target-detector distance of 2.20 m (see Fig.~\ref{setup}). 
        A beam counter H6 (from NA47   
        experiment~\cite{na47}) consisting of plastic
	scintillators is placed 
        directly in the muon beam
	and is thus exposed to rates of about 20 MHz, during the
	two-seconds beam-on phase. To reduce the detector rate to a
        manageable level, the beam counter is segmented
	into 4 rings, each divided in 16 counters thus reducing the
	rate per phototube to at most 1 MHz in
	the center. Several large plastic scintillators are placed
	upstream from the neutron detectors and are used to veto muons
	originating from the beam halo. An iron hadron absorber was
	placed far behind the target along the beam axis.
\subsection{The neutron detectors and electronics}
        Each detector is a 20 x 20 cm cylindrical glass vessel,
        equipped with two photomultipliers (Philips XP4512
        photomultipliers with fast 1.1 nsec RMS pulse) and filled
        with Bicron 501A liquid scintillator which possesses excellent
        pulse shape discrimination properties for neutron
        identification~\cite{501A}. Thin plastic
	scintillators (7 mm) are positioned in front of and around
        each neutron detector to tag charged particles, such as 
        protons and pions. They are also
        used to veto the cosmic muons. 

        The particle energy is measured
        via the time-of-flight (TOF) from its production site in the
        target. The signal of the neutron detector provides the TOF
        start, and the signal of the segmented beam counter, 
        located 6 m upstream
        from the target, is delayed to provide the TOF stop. 
        The analog signals of each of the neutron counters N 
        are coded using 3 different CAMAC ADC's digitizing
        respectively the full analog pulse, the pulse rise and the
        pulse tail. The pulse itself provides the three ADC gates by
        means of a low-threshold discriminator, which also provides a
        CAMAC multi-hit TDC with the TOF start signal. The analog
        signals of each of the plastic scintillators S 
        are coded
        using 15 different CAMAC ADC's.
\section{Neutron detector efficiency}
        \label{nde}  
        The neutron detection efficiency was simulated using the
        GCALOR Monte Carlo package~\cite{gcalor}. Neutrons were generated with
        energies from 2 to 1000 MeV. The efficiency for neutrons entering
        the detector was obtained as a function of both the detector
        energy threshold and the neutron energy.
        The geometrical acceptance of the detector, the quenching effect
        and each detector threshold were taken into account. 
        The calculation of
        scintillator quenching was taken from Ref.~\cite{cecil}, 
        as the amount of light
        depends on the particle type and is not directly proportional to the
        deposited energy. Above a neutron energy of 50 MeV, the
        electron equivalent energy in the scintillator is assumed to
        be a linear function of 
        the proton energy~\cite{quench1}, with a proportionality
        constant of 0.83. Below 50 MeV, we applied
        the following formula:
        \begin{equation}
        \label{cec}
        T_e\ =\ 0.83\  T_p \ -\ 2.82\ \left[\ 1.0-e\ ^{(\ -\ 0.25\ {T_p^{0.93}})}\right]
        \end{equation}
        where $T_e$ was the electron kinetic energy in MeV and $T_p$ the
        proton kinetic energy in MeV.

\begin{figure}
\begin{center}
\includegraphics*[width=10cm]{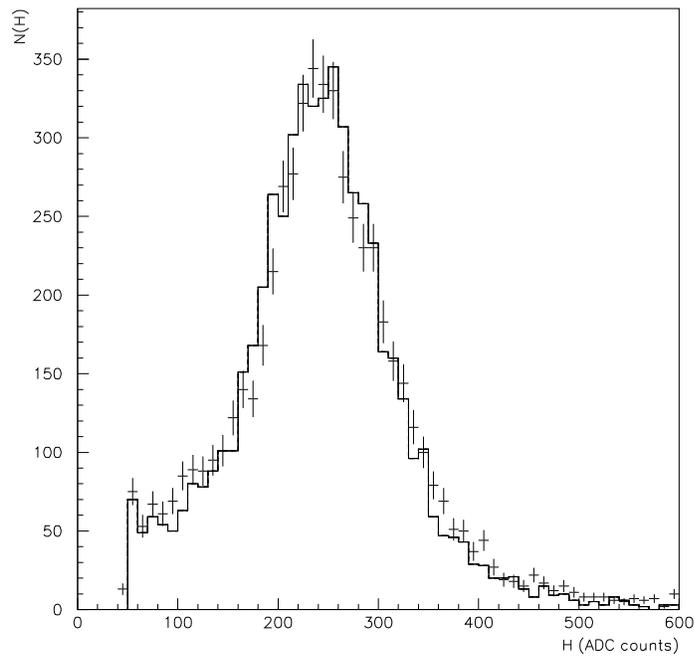}
\vskip 1cm
\caption{\label{muons_guy}  Measured (error bars) and simulated
(histogram) energy
spectra of cosmic muons for the N3 detector. The lower energy
threshold, on the left, is clearly visible and well determined within
2 channels and corresponds to $6.9 \pm 0.3$ MeV electron equivalent
energy.}
\end{center}
\end{figure}

        As will be shown later, the neutron energy spectrum in each
        counter is peaked at low energies.  Therefore, the detection
        efficiency is strongly dependent upon the neutron detectors
        thresholds.  The thresholds are determined from the energy
        loss spectrum of cosmic muons in the scintillator cells.  Vertical
        muons (cosmics) are selected by requiring a coincidence signal
        above pedestals in the top and bottom scintillation counters
        during beam-off runs. Because of the hardware thresholds,
        these spectra show a clear cut-off at low energy. The neutron
        detector thresholds are then extracted by comparing a
        simulated energy spectrum of cosmic muons with the measured
        one. The Monte Carlo spectrum of cosmic muons $N(E)$ is then
        convoluted with a gaussian response function, $R(E,H)$ for
        the detectors:
        \begin{equation}
                N(H) = \int_{0}^{E} R(H,E)N(E)dE
        \end{equation}

        where $N(H)$ is the measured spectrum in ADC counts $H$. The response
        function parameters are of the same form as the ones used
        by Arneodo {\em et al.}~\cite{501A}
        who have described and tested the response of similar
        Bicron 501A liquid scintillator cells. Applying this
        procedure, we obtain thresholds in electron equivalent energy
        of $4.8 \pm 0.4$ MeV for N1, $5.5 \pm 0.3$ MeV for N2 and
        $6.9 \pm 0.3$ MeV for N3. In neutron or proton recoil energy they
	are 8.7$\pm$0.7 MeV, 9.6$\pm$0.5 MeV and 11.4$\pm$0.5 MeV for N1, N2 
	and N3 respectively.
	Fig.~\ref{muons_guy} shows a
        comparison of the measured and simulated, $N(H)$,
        energy spectra of cosmic muons for the N3 detector.
        The errors on the thresholds are assigned to be 2 channels
        (see Fig.~\ref{muons_guy}) and are overestimated.

\begin{figure}
\begin{center}
\includegraphics*[width=12cm]{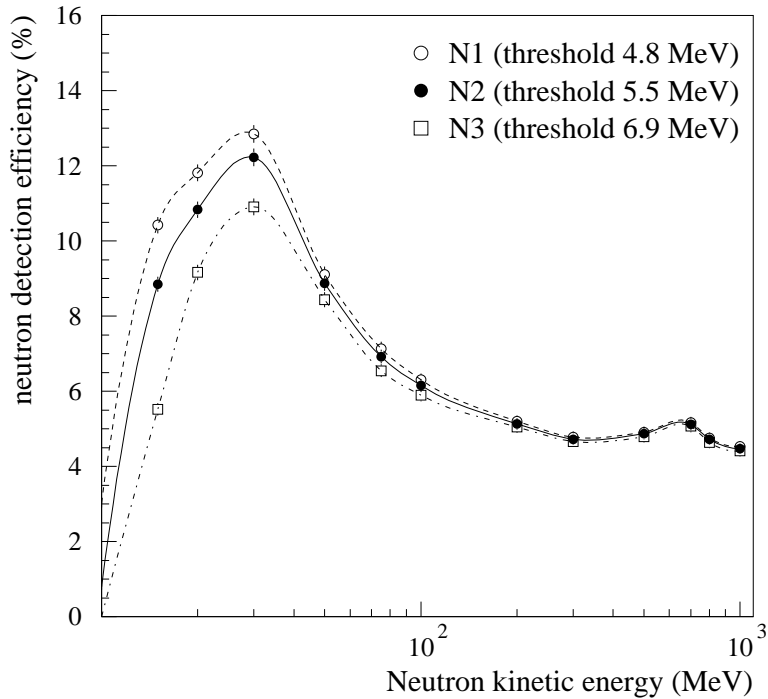}
\caption{\label{quench}Simulation of the neutron detection efficiency as a function
of neutron energy and threshold (in electron equivalent energy), taking 
into account the geometrical
acceptance of the detector and the quenching effect, for the angles
45$^\circ$ (N1), 90$^\circ$ (N2) and 135$^\circ$ (N3). Polynomial
interpolations (dotted, solid and dashed lines) have been obtained
from the simulated efficiencies (open circles, solid circles and open
squares) and used for the efficiency corrections.
The bump at 700 MeV is the result
of an approximation.} 
\end{center}
\end{figure}

        Finally, Fig.~\ref{quench} shows
        the Monte Carlo simulation of the neutron
        detection efficiency as a function of both neutron energy and
        energy threshold. 
        Neutrons 
        were assumed to be detected only if recoil
        protons ionize heavily and stop in the scintillator producing
        delayed light and thus making pulse shape discrimination
        possible. 
\section{Data analysis}
        The results of the analysis of our data from the different
        targets is presented below.
\subsection{Background}
        The challenge of this experiment is to discriminate neutrons
        from the abundant \linebreak bremsstrahlung
        photons. Furthermore, four other background sources may contribute
        to the measured spectra: charged particles from the
        target, cosmic muons, muons from the beam and ambient
        neutrons. Cosmic muons and
        charged particles can be rejected using the plastic
        scintillators. Muons from the beam halo
        are eliminated by the dedicated veto counter (HV, see
        Fig.~\ref{setup}). Surrounding
        neutrons cannot be eliminated, but their contribution can be
        estimated from dedicated empty-target runs. Finally, neutrons and
        photons are identified by a pulse shape discrimination (PSD) of
        their signal in the liquid scintillator.
\subsection{Pulse shape discrimination}
        \label{psd}
\begin{figure}
\begin{minipage}[t]{68mm}
\includegraphics*[width=7.5cm]{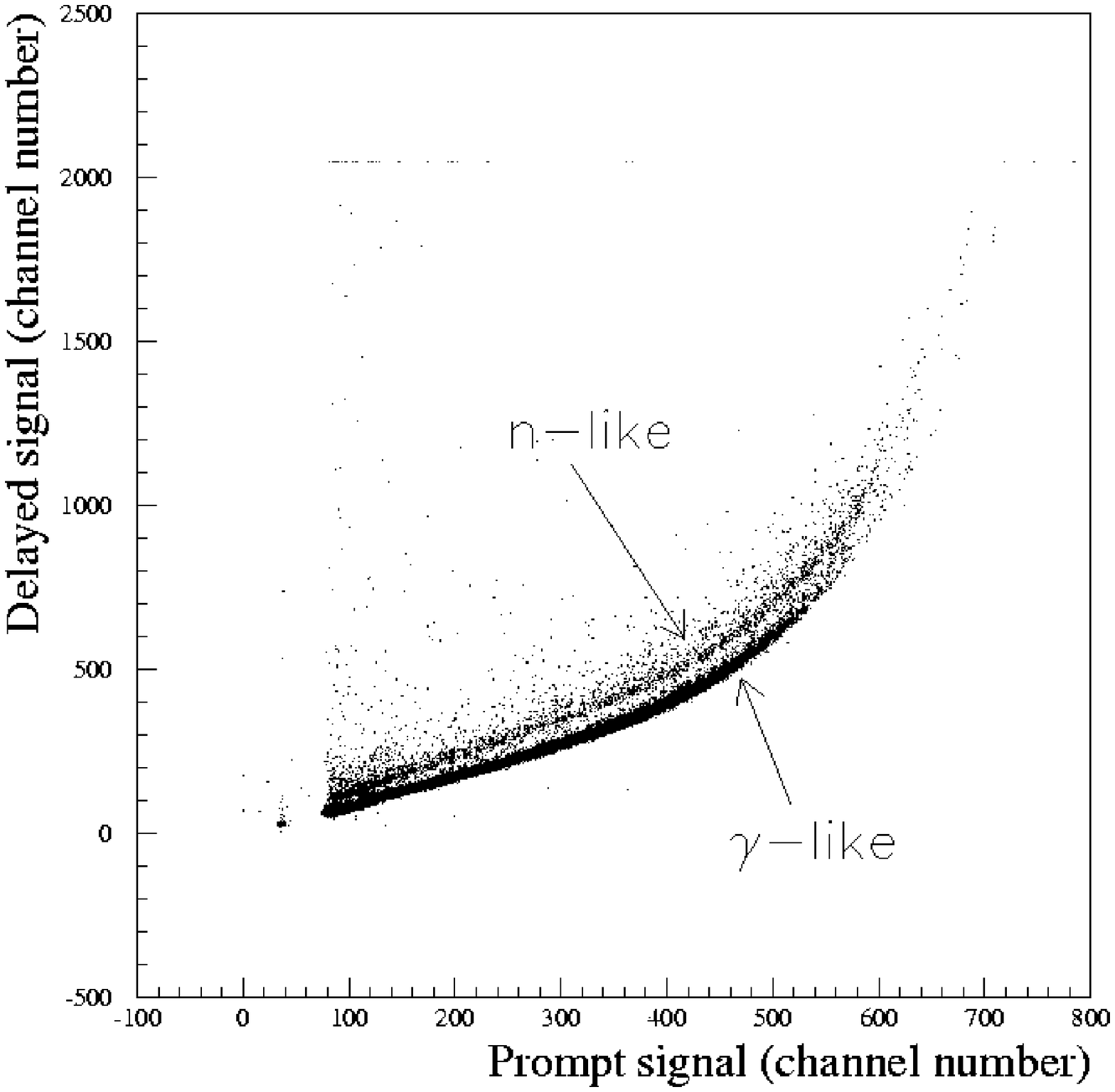}
\caption{\label{discri1}First pulse shape discrimination (PSD1) for
the N1 detector and the carbon target. The
signal is measured in two different time intervals : prompt signal
amplitude vs. delayed signal amplitude.}
\end{minipage}
%
\hspace*{1.cm}
\begin{minipage}[t]{68mm}
\includegraphics[width=7.5cm]{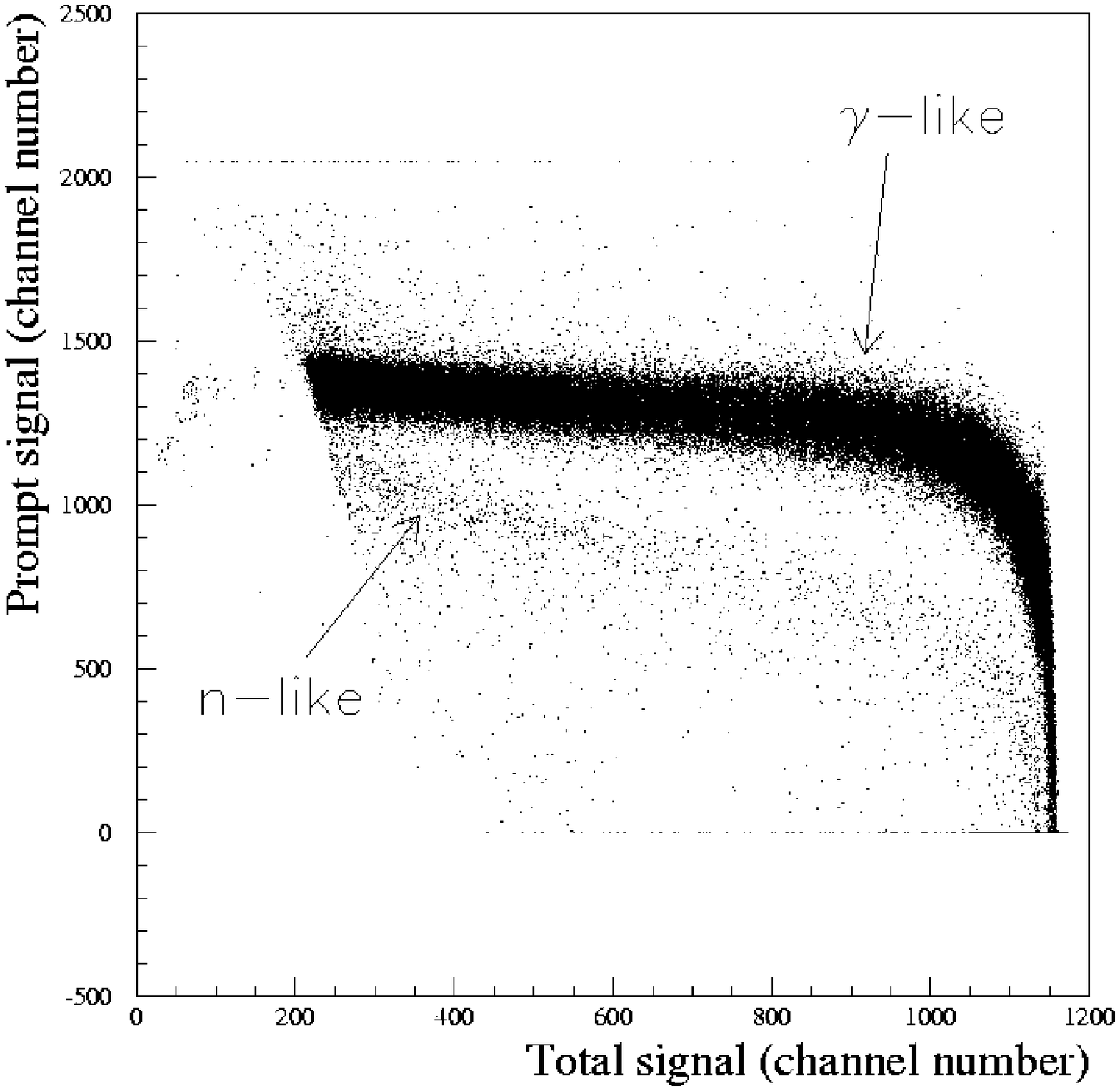}
\caption{\label{discri2} Second pulse shape discrimination (PSD2) for
the N1 detector and the carbon target. Prompt
signal amplitude vs. total signal amplitude (different binning).}
\end{minipage}
\end{figure}

\begin{figure}
\begin{center}
\includegraphics*[width=9cm]{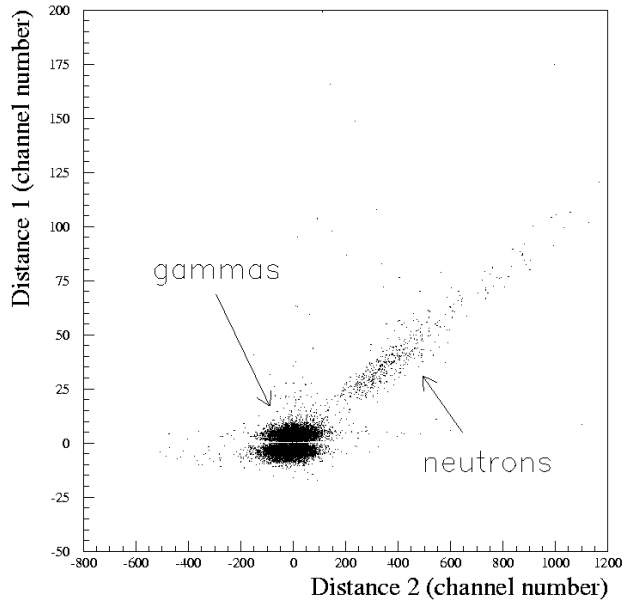}
\vskip 1cm
\caption{\label{dist}Distance~1 vs. distance~2 for the N1 detector and
the carbon target. 
These distances were
calculated between every point and the $\gamma$-like events 
for PSD1 and PSD2. Only neutral particles are selected.}
\end{center}
\end{figure}

        The time-development of scintillation light depends on the
        nature of the ionising particle. In scintillators such as 
	Bicron 501A higher ionisation density due to slow charged particles, 
	such as recoil protons from neutron scattering,
	leads to the excitation
	of longer lived states and thus to a slower component in the 
	light pulse. Measuring the
        signal amplitude at 2 different time intervals allows to visualize
	this in a PSD
        plot~\cite{psd1}. The complete PSD method is explained in
        Ref.~\cite{knoll}. Fig.~\ref{discri1} shows the amplitude 
        obtained in a delayed  window
        as a function of the amplitude  measured in a prompt one (PSD1).
        Based on the same principle,
        Fig.~\ref{discri2} is obtained by comparing the amplitude in the
        prompt window with the total signal amplitude
        (PSD2). In
        these two different PSD representations, we can distinguish a
        $\gamma$-like curve containing bremsstrahlung photons and
        cosmic and remaining beam halo muons,
        and a neutron-like zone containing
        all the neutrons and the charged hadrons from the target.
        The $\gamma$-like events are easily selected from the
        time-of-flight since they are prompt with respect to the beam
        muons. Then we calculate for each event the
        real distance (in units of channel number) 
        between the corresponding amplitudes and the fitted 
        $\gamma$-like curve for PSD1 (distance~1)
        and PSD2 (distance~2). We can then represent distance~1 as
        a function of distance~2, for all events including the charged
        particles. Figure~\ref{dist} shows 
        the events left after the cuts which
        select only neutral particles. A clear separation can be seen
        between photons and neutrons. 
        A cut is applied to select neutron like events
        achieving nearly complete suppression of the 
        $\gamma$-like events. The acceptance is calculated by
	applying a gaussian fit on the projected
	neutron and gamma peaks, and evaluating the relative area
	under the gamma peak above threshold.
        The acceptance was 
	calculated separately for each detector and target. This was 
	necessary because of the limited stability of the PSD, and because
	of the differences in the relative strengths of the neutron and gamma
	peaks from case to case. The global 
	cut acceptances, reflecting these variations between
	targets and detectors, are shown in Table \ref{ta:ctacp}. 
	The accepted neutron events are then
        used for the time-of-flight analysis.
\subsection{Time-of-flight analysis}
        \label{tof}
\begin{figure}
\begin{center}
\includegraphics*[width=10cm]{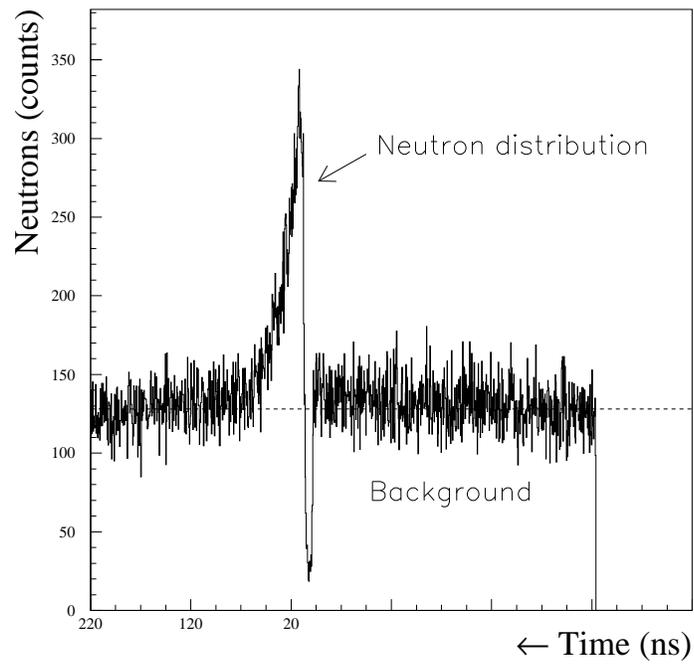}
\vskip 1cm
\caption{\label{time}TDC spectrum corresponding to the neutron 
distribution dn/dt for the carbon target. The undershoot for times immediately before the
'prompt' signal is caused by the removal of the gamma peak. The
background due to uncorrelated beam muons is flat.}
\end{center}
\end{figure}

        Before being converted into energy distributions, the selected
        neutron time-of-flight spectra must be corrected for
        off-target events and for random TOF stop signals. The first
        background arises from triggers correlated with beam
        particles, as neutrons bouncing on the ground or walls, for
        example. Their contribution can be deduced from 'empty target'
        runs properly normalized to the incident flux. The second
        background component is due to multiple TOF stop signals given
        by randomly incoming muons hitting the H6 detector
        scintillators. As their time distribution is flat (see 
        Fig.~\ref{time}), they can be
        easily subtracted.

\begin{figure}
\begin{center}
\includegraphics*[width=15cm]{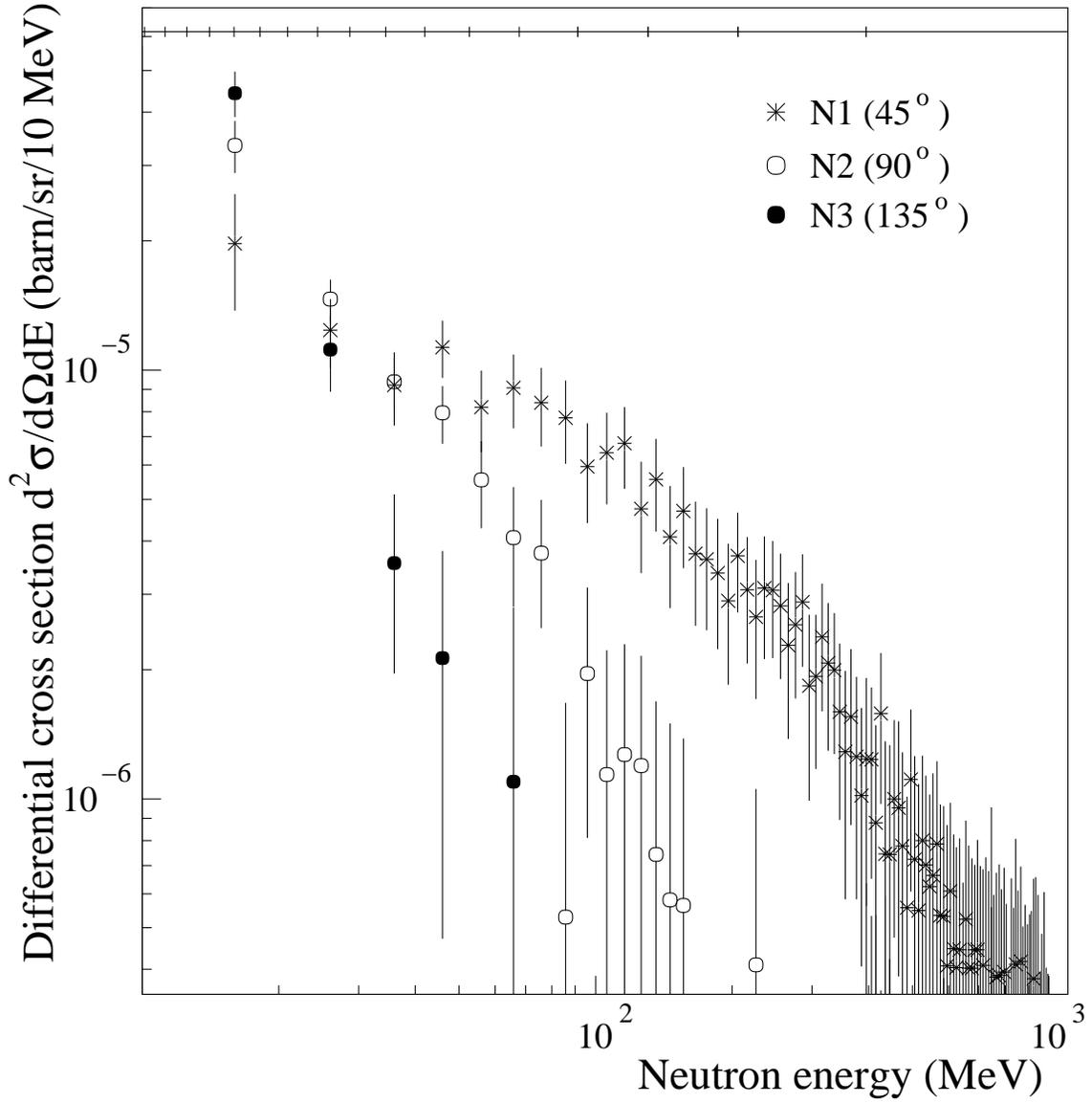}
\caption{\label{energiec}Neutron energy spectrum for the carbon target, 
including all the
corrections. The spacing between two ticker marks on the upper 
horizontal line
gives the energy resolution (FWHM) at the corresponding energy, 
derived from the time resolution, for the detector N1.}
\end{center}
\end{figure}

        The calibration of the time spectrum is obtained from the
        gamma peak. The time resolution slightly depends on the
        target. As an example, it is 1.0 ns, 4.6 ns and 3.8 ns for
        N1, N2 and N3 respectively for the carbon target. In this case
        we measure
        maximum neutron energies of around 1 GeV, 250 MeV and 310 MeV
        for N1, N2 and N3 respectively. The time resolution determines the
	energy resolution, which is shown as function of energy 
	in Fig. \ref{energiec}.
        The energy distribution is computed from the TOF distribution,
        and corrected
        for the detector efficiency as described in section~\ref{nde}.
\section{Results}
\subsection{Additional acceptance corrections}
        The charged particle rejection using the scintillators S 
        leads to a loss of good neutron events. The corresponding
        acceptances are determined from the comparison of neutron
        samples with and without scintillator cut for each neutron
        detector, and strongly depend on the scintillator
        efficiencies and thresholds. In
        addition, the 85 Hz trigger rate for all detectors above a
        threshold of 50 MeV leads to a dead time of 2.1$\%$. All the
        efficiencies taken into account in the calculation of the
        neutron production cross section are summarized in Table~\ref{eff}.
\subsection{Production cross section and energy spectrum}
\begin{figure}
\begin{center}
\includegraphics*[width=15cm]{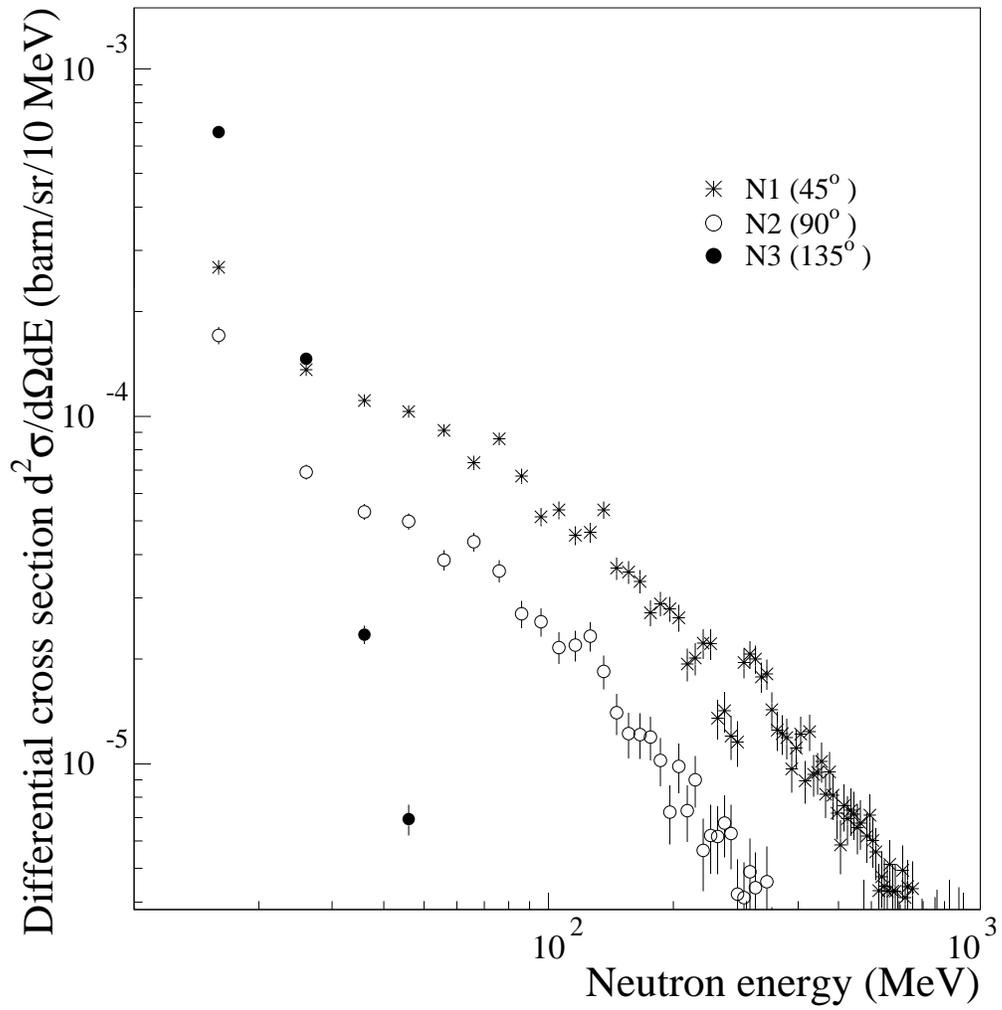}
\caption{\label{energiecu}Neutron energy spectrum for the copper target, including all the
corrections.}
\end{center}
\end{figure}

\begin{figure}
\begin{center}
\includegraphics*[width=15cm]{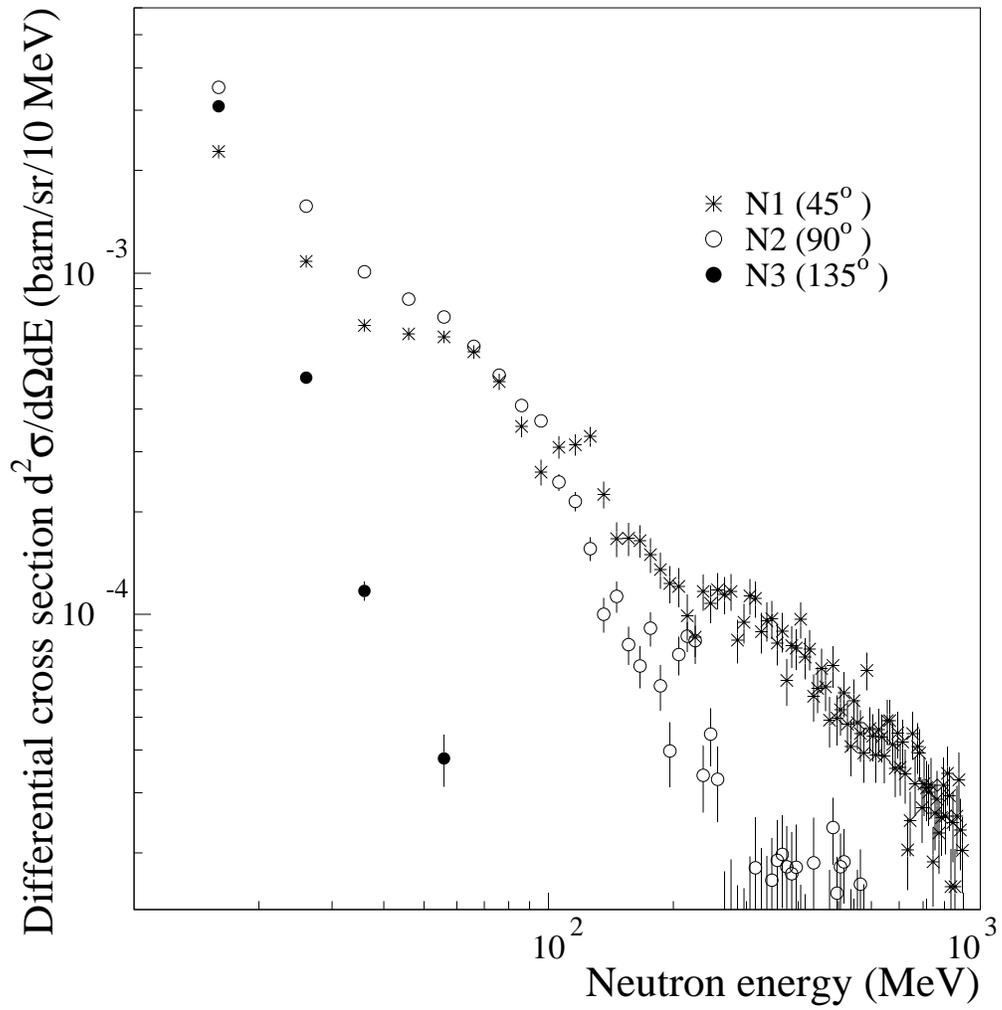}
\caption{\label{energiepb}Neutron energy spectrum for the lead target, including all the
corrections.}
\end{center}
\end{figure}

        The neutron production cross section was calculated using the
        following formula:
        
        \begin{equation}
        \label{sigma}
        {d\sigma\over{d\Omega}}(\Theta\pm\Delta\Theta)=\left({dn\over{\phi\ N\ d\Omega}}\right)
        \end{equation}

        where dn is the number of selected neutrons corrected for all
        acceptances.  
        All parameters are given in Table~\ref{res}. 
        Figures~\ref{energiec}, \ref{energiecu} and~\ref{energiepb}
        show the differential neutron production cross
        section as a function of the neutron energy, including all
        efficiency corrections, for the three angles, and
        for the different targets. The
        energy integrated values from the corresponding threshold up, 
        are gathered in Table 3. It is worth
        noting that the neutron energy spectrum is displaced towards
        low energies when going from forward to backward angles. As a
        consequence, no neutrons are detected above 70 MeV at
        135$^\circ$. 
        Our measurements show that
        the angular distribution is forward peaked, with the exception
        of the lowest energy bin. This effect can be understood as due
        to the recoil of the source nucleus combined with an isotropic
        distribution of the emitted neutron at the lowest energies.
        One also observes that the angular distribution gets wider 
	with increasing atomic number. Generally our angular and energy
        distributions agree
        with the FLUKA simulation performed
        by Wang {\em et al.}~\cite{fluka}. The slope of the neutron 
	energy spectrum for carbon is in a qualitative
	agreement with the slope $\sim E^{-1}$ (integrated over 
	all angles) found
	in Ref. \cite{LVD_n}.

\begin{figure}
\begin{center}
\includegraphics*[width=15cm]{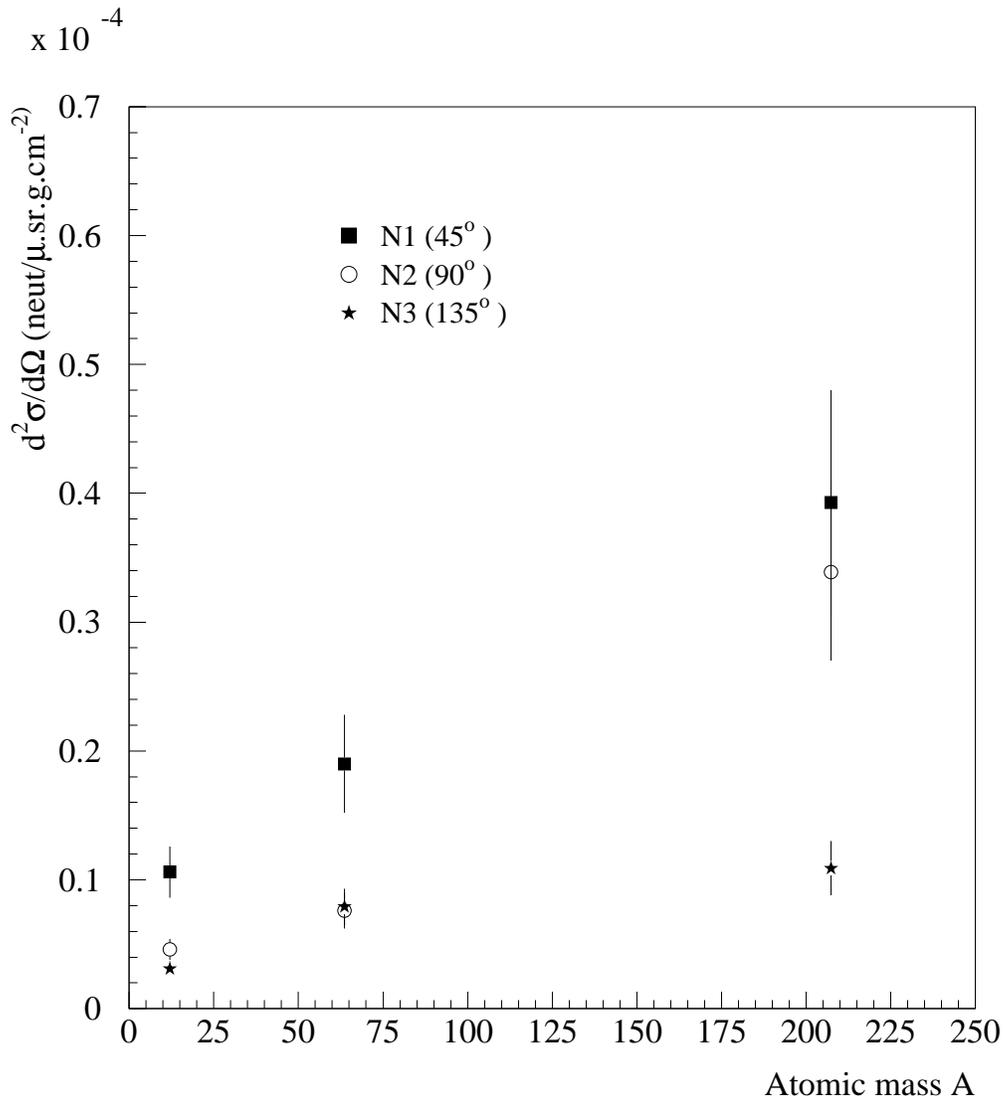}
\caption{\label{seceff}Integrated cross section vs atomic number for
the three different angles.}
\end{center}
\end{figure}

	Figure~\ref{seceff} shows the integrated cross section as a
        function of the atomic number for the three different
        angles and three different targets. We note that the cross 
	section increases significantly
	with the
        atomic number.
\section{Conclusion}
        We investigated the production of spallation neutrons obtained
        from 190~GeV/c muons scattering on graphite, 
        copper and lead targets. 
        The neutrons were observed by liquid scintillator
        detectors, allowing background rejection by means of pulse
        shape discrimination. The neutron energy distribution was
        determined via time-of-flight. 
        The 190 GeV muon energy corresponds to the
        mean energy of cosmic-ray muons at underground experimental
        sites of about 2000 meters
        water equivalent depth. 

	It should be noted, however, that
	in the present experiment only 
        neutrons associated with the primary muon-nuclear spallation 
        process are detected, while in the underground detectors 
        the total neutron yields are measured. 
        Thus the present experiment is not dependent on the
        subsequent neutron transport and multiplication. In view of this,
        and also in the view of the difference in the muon spectrum
        (narrow vs. broad distribution) direct comparison is difficult.
        Nevertheless, the cross sections quoted in 
        Table 3 are in a qualitative
        agreement with the yields underground and, as mentioned above,
        their shape agrees with the FLUKA simulation.

        Results on neutron angular distribution and
        energy distribution were obtained for the first time. 
        The differential cross section as a
        function of the neutron energy were obtained for 45$^\circ$,
        90$^\circ$ and 135$^\circ$ production angles with around
        15$\%$ accuracy.
\section*{Acknowledgements}
        Special thanks go to G. Gervasio for his contribution to the
        data decoding and O. Drapier for many useful
        discussions.  We are grateful to
        D. Hilscher of HMI, Berlin for providing us with a PSD
        module. We thank H. Wong, A. Schopper, L. Gatignon and the staff
        at CERN for assistance provided.
        This project was supported in part by the US
        Department of Energy.
%

\vfill\eject

\hspace*{-0.5cm}

\begin{table}
\vspace*{1.cm}
\caption{\label{eff}Acceptance of the cuts \label{ta:ctacp}}
\begin{tabular}{l l c c c}
\hline
Target & Analysis cuts & N1 (45$^\circ$) & N2 (90$^\circ$) & N3 (135$^\circ$)\\
\hline
Carbon & PSD cuts  & 87$\%$ & 66$\%$ & 64$\%$\\
& Plastic scintillators & 67 $\%$ & 82 $\%$ &
66 $\%$ \\
\hline
Copper & PSD cuts  & 48$\%$ & 91$\%$ & 26$\%$\\
& Plastic scintillators & 69 $\%$ & 85 $\%$ &
88 $\%$ \\
\hline
Lead & PSD cuts  & 50$\%$ & 28$\%$ & 28$\%$\\
& Plastic scintillators & 57 $\%$ & 84 $\%$ &
87 $\%$ \\
\hline
& Life time & \multicolumn{3}{c}{97.9$\%$}\\
\hline
\end{tabular}
\end{table}

\begin{table}
\caption{\label{res}Summary of calculated cross-section parameters}
\begin{tabular}{l c c}
\hline
parameter & target & value\\
\hline
Angular acceptance $d\Omega$  & & (5.2\ $\pm$\ 0.6)\ x\ 
10$^{-4}$ sr\\
\hline
Muons on target $\phi$ & C & 1.63\ x\ 10$^{11}$ muons\\
& Cu & 1.84\ x\ 10$^{11}$ muons\\
& Pb & 1.93\ x\ 10$^{11}$ muons\\
Target density N=$\rho$L & C & (8.52\ $\pm$\ 0.11)\ x\ 10$^{24}$
atoms/cm$^2$\\
& Cu & (2.12\ $\pm$\ 0.03)\ x\ 10$^{24}$ atoms/cm$^2$\\
& Pb & (3.30\ $\pm$\ 0.04)\ x\ 10$^{23}$ atoms/cm$^2$\\
\hline
\end{tabular}
\begin{tabular}{l c c c}
Total number of neutrons, & & 45$^\circ$  & 100500 neutrons\\
above threshold, corrected &C &  90$^\circ$  &  53510 neutrons\\
for the detector efficiency  & &  135$^\circ$  & 29395 neutrons\\
\hline
& & 45$^\circ$  & 277960 neutrons\\
& Cu &   90$^\circ$  &  135545 neutrons\\
 & &  135$^\circ$  & 148955 neutrons\\
\hline
& & 45$^\circ$  & 251360 neutrons\\
& Pb &   90$^\circ$  &  316820 neutrons\\
 & &  135$^\circ$  & 105060 neutrons\\
\hline
\end{tabular}
\end{table}

\begin{table}
\caption{\label{cross}Differential neutron production cross-section}
\begin{tabular}{l c c c}
\hline
Differential cross section & target &barn/sr & neut/$\mu\cdot$sr$\cdot$g$\cdot$cm$^{-2}$\\
\hline
${{d\sigma}\over{d\Omega}}(45^\circ\pm1^\circ)$ &C& (0.21\ $\pm$\ 0.04)\ x\ 
10$^{-3}$ & (1.06\ $\pm$\ 0.20)\ x\ 
10$^{-5}$\\
${{d\sigma}\over{d\Omega}}(90^\circ\pm1^\circ)$ & & (0.09\ $\pm$\ 0.02)\ x\ 
10$^{-3}$ & (0.46\ $\pm$\ 0.08)\ x\ 
10$^{-5}$\\
${{d\sigma}\over{d\Omega}}(135^\circ\pm1^\circ)$ & & (0.06\ $\pm$\ 0.01)\ x\ 
10$^{-3}$ & (0.31\ $\pm$\ 0.06)\ x\ 
10$^{-5}$\\
\hline
${{d\sigma}\over{d\Omega}}(45^\circ\pm1^\circ)$ &Cu& (2.00\ $\pm$\ 0.40)\ x\ 
10$^{-3}$ & (1.90\ $\pm$\ 0.38)\ x\ 
10$^{-5}$\\
${{d\sigma}\over{d\Omega}}(90^\circ\pm1^\circ)$ & & (0.80\ $\pm$\ 0.15)\ x\ 
10$^{-3}$ & (0.76\ $\pm$\ 0.14)\ x\ 
10$^{-5}$\\
${{d\sigma}\over{d\Omega}}(135^\circ\pm1^\circ)$ & & (0.83\ $\pm$\ 0.15)\ x\ 
10$^{-3}$ & (0.79\ $\pm$\ 0.14)\ x\ 
10$^{-5}$\\
\hline
${{d\sigma}\over{d\Omega}}(45^\circ\pm1^\circ)$ &Pb& (13.53\ $\pm$\ 3.00)\ x\ 
10$^{-3}$ & (3.93\ $\pm$\ 0.87)\ x\ 
10$^{-5}$\\
${{d\sigma}\over{d\Omega}}(90^\circ\pm1^\circ)$ & & (11.67\ $\pm$\ 2.38)\ x\ 
10$^{-3}$ & (3.39\ $\pm$\ 0.69)\ x\ 
10$^{-5}$\\
${{d\sigma}\over{d\Omega}}(135^\circ\pm1^\circ)$ & & (3.75\ $\pm$\ 0.72)\ x\ 
10$^{-3}$ & (1.09\ $\pm$\ 0.21)\ x\ 
10$^{-5}$\\
\hline
\end{tabular}
\end{table}

\end{document}